# Davydov's solitons in a homogeneous nucleotide chain.

**Victor D. Lakhno**

*Institute of Mathematical Problems of Biology, RAS, 142290, Pushchino, Institutskaya str., 4, Russia,* lak@impb.psn.ru

**Abstract.** Charge transfer in homogeneous nucleotide chains is modeled on the basis of Holstein Hamiltonian. The path length of Davydov solitons in these chains is being studied. It is shown that in a dispersionless case, when the soliton velocity $V$ is small, the path length grows exponentially as $V$ decreases. In this case the state of a moving soliton is quasisteady. In the presence of dispersion determined by the dependence $\Omega^2 = \Omega_0^2 + V_0^2 \kappa^2$ the path length in the region $0 < V < V_0$ is equal to infinity. In this case the phonon environment follows the charge motion. In the region $V > V_0$ the soliton motion is accompanied by emission of phonons which leads to a finite path length of a soliton. The latter tends to infinity as $V \to V_0 + 0$ and $V \to \infty$. The presence of dissipation leads to a finite soliton path length.

An equilibrium velocity of soliton in an external electric field is calculated. It is shown that there is a maximum intensity of an electric field at which a steady motion of a soliton is possible. The soliton mobility is calculated for the stable or ohmic brunch.

## 1.    Introduction

An idea of modern molecular nanoelectronics is to use molecular chains as wires [1–4]. This highlights the problem of conducting properties of such molecular wires. In the pioneering works by Davydov [5-10] it was shown that the main carriers of excitation in molecular wires are solitons (or polarons). Though to date an extensive literature has been accumulated on the subject, it is a gross exaggeration to say that everything is absolutely clear now.

Let us consider a molecular chain consisting of biatomic molecules in which atoms can displace from their equilibrium positions. This approximation is known as a Holstein model [11] which represents one of the simplest models of a deformable chain. It is widely used in describing charge transfer in DNA where the role of biatomic molecules is played by Watson-Crick nucleotide pairs [11-15]. Notwithstanding its simplicity the model does not have an analytical solution.

For the first time, the points to be clarified in the context of this problem were brought to notice by Landau [16]. According to Landau, even at zero temperature, an electron moving over such a chain can cause the chain's deformation which can catch the electron to form a localized state. In Pekar's works this localized state was called a polaron [17]. Polaron states have been studied for more than half a century and still remain to be understood.

One of the questions to be answered was that of whether a polaron state can move over the chain. It would seem, an unequivocal and principal answer was given in the paper by Landau and Pekar [18]. According to Landau and Pekar, the case of a deformable chain differs from the case of a rigid one in that in the former case a polaron moves as a whole while in the latter one an electron does.

This conclusion was doubted by Davydov and Enolskii [19-21]. They paid attention to the fact that in the case of optical phonons their group velocity is equal to zero and, in the absence of dispersion, the deformation induced by the electron cannot follow it. According to Davydov and Enolskii, a steady motion of solitons over a chain is possible only if the chain dispersion is taken into account.

From these considerations, the authors of [22-25] concluded that in a dispersionless case, any arbitrarily weak interaction such as, for example, an external electric field, which would force the electron to move, would lead to its destruction and transformation into Bloch state.

The conclusion of the instability of polaron states in response to weak external influences is in contradiction with the fact that a polaron state is energetically more advantageous than delocalized Bloch states.

As will be shown in this work, this contradiction results from the fact that the state of a uniformly moving polaron, considered by Landau and Pekar is, actually, not the eigen state of the initial Hamiltonian. When moving along the chain, the electron will inevitably actuate the chain's atoms, leaving behind a "tail" of oscillating molecules. This will lead not to destruction of the polaron, but to its stopping. Therefore the question arises of how suitable is the concept of moving polaron states in describing the transfer processes in polynucleotide chains.

The paper is arranged as follows.

In Section 2 we introduce a semiclassical Holstein model for a molecular chain. An expression is obtained for the Green function of the classical equation of the chain motion with regard for dissipation. In the absence of dissipation it transforms into the equation obtained by Davydov and Enolskii in [19].

In Section 3 we show that in a molecular chain with dispersionless phonons an excess electron can be in a "quasisteady" moving soliton state, the electron path length tending to infinity exponentially as the soliton velocity tends to zero. This fact justifies the idea of the possibility of "quasisteady" states of solitons in the case of small velocities. It is shown that when the soliton moves at a finite velocity, its stopping is caused by the oscillation excitations behind it.

In Section 4 we show that when dispersion is taken into account, the picture considered changes qualitatively. According to Davydov's results, in this case as $V < V_0$, the soliton does not emit phonons and represents the eigen state of the Hamiltonian. The soliton path length then goes into infinity.

In Section 5 we consider the soliton motion in the presence of dissipation. An expression is obtained for the soliton mobility in the absence of dispersion at zero temperature.

In Section 6 we investigate the steady motion of a soliton in an electric field. It is shown that there exists a maximum value of the electric field intensity at which the steady motion is possible. A general expression is obtained for the dependence of the soliton equilibrium velocity on the field intensity. The dependence contains two branches, one of which is stable ohmic, and the other is unstable nonohmic.

In Section 7 we obtain an expression for the mobility of moving soliton with regard for dispersion for T=0.

In Section 8 we apply the obtained results for homogeneous polynucleotide chains.

In Section 9 we discuss the results obtained.

## 2. Motion equations of Davydov's soliton in the continuum approximation.

Following [19], let us write down Hamiltonian H describing the electron motion in a molecular chain in the continuum approximation as:

$$H = H_e + H_{int} + H_{ph},$$

(1)

$$H_e = -\frac{1}{a}\int \Psi^*(x,t)\frac{\hbar^2}{2m}\frac{\partial^2}{\partial x^2}\Psi(x,t)\,dx, \tag{2}$$

$$H_{int} = \frac{\chi}{a}\int |\Psi(x,t)|^2 u(x,t)\,dx, \tag{3}$$

$$H_{ph} = \frac{M}{2a}\int\left[\left(\frac{\partial u(x,t)}{\partial t}\right)^2 + \Omega_o^2 u^2(x,t) + V_0^2\left(\frac{\partial u(x,t)}{\partial x}\right)^2\right]dx. \tag{4}$$

In (1) – (4) the energy is reckoned from the bottom of the conductivity band, $\Psi(x,t)$ normalized electron wave function:

$$\frac{1}{a}\int |\Psi(x,t)|^2\,dx = 1 \tag{5}$$

where $a$ is the lattice constant. Unless otherwise specified, integration in (1) – (5) and all the ensuing expressions is carried out in the infinite limit. The quantity $m$ is the electron effective mass, $\chi$ is the constant for the electron interaction with the chain displacements $u(x,t)$, $M$ is the reduced mass of an elementary cell, $\Omega_0$ is the frequency of intramolecular oscillations of the chain, $V_0$ has the meaning of the minimum phase, or maximum group velocity of the chain oscillations. Notice that Hamiltonian $\hat{H}$ is a continuum analog of the discrete model descriptive of an excess electron inserted in a homogeneous polynucleotide chain [12–15].

Motion equations corresponding to Hamiltonian (1) – (4) have the form:

$$\frac{\partial^2 u}{\partial t^2} + \Omega'\frac{\partial u}{\partial t} + \Omega_0^2 u - V_0^2\frac{\partial^2 u}{\partial x^2} + \frac{\chi}{M}|\Psi|^2 = 0, \tag{6}$$

$$i\hbar\frac{\partial \Psi}{\partial t} + \frac{\hbar^2}{2m}\frac{\partial^2 \Psi}{\partial x^2} - \chi u\,\Psi = 0, \tag{7}$$

The term $\Omega'\partial u/\partial t$ in the left-hand side of (6) describes oscillations attenuation due to friction, $\Omega'$ is the frequency of the oscillations damping.

For the steady solutions of the form:

$$u(x,t) = u(\xi), \qquad \Psi(x,t) = \varphi(\xi)\exp\frac{i}{\hbar}\left[mVx - \left(W + \frac{1}{2}mV^2\right)t\right],$$

$$\xi = (x - Vt)/a, \tag{8}$$

equations (6), (7) are written as:

$$-\varepsilon\Omega_0^2\frac{d^2u}{d\xi^2} - 2\Omega_0^2\delta\frac{du}{d\xi} + \Omega_0^2 u + \frac{\chi}{M}|\varphi|^2 = 0, \tag{9}$$

$$-\frac{\hbar^2}{2ma^2}\frac{d^2\varphi}{d\xi^2}+\chi u\varphi=W\varphi, \tag{10}$$

where:

$$\delta=\frac{\Omega'V}{2\Omega_0^2 a}, \qquad \varepsilon=\left(V_0^2-V^2\right)/a^2\Omega_0^2 \tag{11}$$

Equation (9) is solved with the use of the Fourier transform. As a result $u(\xi)$ will take the form:

$$u(\xi)=-\frac{\chi}{M\Omega_0^2}\int d\xi'\omega(\xi'-\xi)|\varphi(\xi')|^2 \tag{12}$$

$$\omega(\xi)=\frac{1}{2\pi}\int dq\,\frac{e^{iq\xi}}{\varepsilon q^2+2i\delta q+1} \tag{13}$$

Calculation of (13) yields:

$$\omega(\xi)=\frac{\Theta(\xi)}{\sqrt{|\varepsilon|-\delta^2}}e^{-\frac{\delta}{|\varepsilon|}\xi}\sin\frac{\xi}{|\varepsilon|}\sqrt{|\varepsilon|-\delta^2} \tag{14}$$

if $\varepsilon<-\delta^2$;

$$\omega(\xi)=\frac{2\pi\Theta(\xi)}{\sqrt{\delta^2-|\varepsilon|}}e^{-\frac{\delta}{|\varepsilon|}\xi}\text{ch}\left(\frac{\xi}{|\varepsilon|}\sqrt{\delta^2-|\varepsilon|}\right) \tag{15}$$

if $-\delta^2<\varepsilon<0$;

$$\omega(\xi)=\delta(\xi)+i\,\Theta(\xi)e^{-\xi/2\delta} \tag{16}$$

if $\varepsilon=0$;

$$\omega(\xi)=\frac{1}{2\sqrt{\delta^2+|\varepsilon|}}\left\{\Theta(\xi)e^{-\frac{\xi}{|\varepsilon|}\left(\sqrt{\delta^2+|\varepsilon|}-\delta\right)}+\Theta(-\xi)e^{\frac{\xi}{|\varepsilon|}\left(\sqrt{\delta^2+|\varepsilon|}+\delta\right)}\right\} \tag{17}$$

if $\varepsilon>0$, where $\delta(\xi)$ is the Dirac $\delta$-function, $\Theta(\xi)=1$, if $\xi>0$, $\Theta(\xi)=0$, if $\xi<0$. For $\delta=0$, i.e. in the absence of dissipation, these expressions go over into those obtained in [19]. The sign of the argument of $\Theta$-function in (14) differs from the sign of the corresponding argument in [19] due to the fact that as $\varepsilon<0$ the complex summand in the denominator of integrand (13) has a different sign.

## 3. Soliton path length in the absence of dispersion and dissipation.

The authors of [19-21] reasoned that in a molecular chain without dispersion a steady motion of a soliton formed by an excess electron is impossible. The arguments were based on the fact that phonons obeying the dispersion law $\Omega^2 = \Omega_0^2 + V_0^2 k^2$ have a zero group velocity as $V_0 = 0$ and therefore the phonon environment cannot follow the soliton motion.

Of interest here is to find the length of a path that a soliton having at the initial moment the velocity $V$ travels until it finally stops. In this section we will deal with the case of the absence of dissipation ($\delta = 0$) and absence of dispersion ($V_0 = 0$).

Using (14) we express $\omega(\xi)$ as:

$$\omega(\xi) = \frac{\Theta(\varepsilon)}{\sqrt{|\varepsilon|}} \sin \frac{\xi}{\sqrt{|\varepsilon|}}, \qquad \varepsilon < 0 \tag{18}$$

As $\varepsilon = 0$, equation (10) is reduced to a stationary nonlinear Schröedinger equation, the normalized solution of which has the form:

$$\varphi(\xi) = \frac{1}{\sqrt{2r}} \operatorname{ch}^{-1}(\xi/r), \qquad r = 4M(\hbar\,\Omega_0)^2/m\chi^2 a^2 \tag{19}$$

In the absence of dispersion and small soliton velocity, when $|\varepsilon| \ll 1$, from (12), (18) we get:

$$u(\xi) = c \sin\left(\xi/\sqrt{|\varepsilon|}\right), \qquad \xi < c_1 r \tag{20}$$

$$c = -\pi r \chi / 2M\Omega_0^2 a |\varepsilon| \operatorname{sh}\left(\pi r / 2\sqrt{|\varepsilon|}\right) \tag{21}$$

where $c_1$ is a constant of the order of 1.

From (20) follows that for $\xi = (x - Vt)/a$ a soliton moving along a path in positive direction leaves behind a "tail" of the chain oscillations loosing on the way its kinetic energy $E_{kin} = m^{**}V^2/2$ where

$$m^{**} = \frac{1}{30} \frac{m^3 a^4}{\hbar^6} \frac{\chi^8}{\Omega_0^{10} M^4} \tag{22}$$

is the soliton effective mass, until it finally stops.

The distance L which a soliton will travel is found from the condition:

$$E_{ph} = E_{kin} \tag{23}$$

$$E_{ph} = \frac{M}{2} \int\limits_{-L/a}^{c_1 r} \left[ \frac{\left(V^2 + V_0^2\right)}{a^2} \left(\frac{du}{d\xi}\right)^2 + \Omega_0^2 u^2 \right] d\xi \tag{24}$$

Substituting expression (20) into (24) and using (23) for small $\varepsilon$ we express the soliton path length as:

$$\frac{L}{a} = \frac{1}{120\,\pi^2} \frac{m^5 \chi^{10} a^4 V^6}{M^5 \Omega_0^{16} \hbar^{10}} \, \mathrm{sh}^2\left(\frac{\pi\, r\, a\, \Omega_0}{2V}\right) \qquad (25)$$

The quantity $L$, thus defined, is the distance at which soliton velocity changes significantly.

So, in the absence of dispersion, as $V \to 0$, the soliton path length $L/a$ tends to infinity $\sim \exp (\pi\, r\, a\, \Omega_0 / V)$ This fact justifies the idea of the possibility of "quasisteady" states of solitons and polarons at small $V$, for which all the calculations of the effective mass were made.

The physical reason why the emission of phonons by a moving soliton is exponentially small is that, according to (13), in the absence of dissipation, contribution into the emission is made by $q$ satisfying the Cherenkov condition:

$$qV = \Omega_0\, a$$

As $V \to 0$ this condition is fulfilled for $q \to \infty$. For a localized state, in this limit case, the Fourier component of the electron density is exponentially small. So, in this limit case only an exponentially small portion of the electron distribution will take part in the phonon emission.

## 4. Soliton path length in the presence of dispersion and absence of dissipation.

To have the general pattern of the dependence of the soliton path length on the velocity $V$ in the case on nonzero dispersion ($V_0 \neq 0$) let us consider some limit cases.

**$\varepsilon < 0$.** When $\varepsilon < 0$, the distribution of displacements over the chain is determined by (18). Since in this case the soliton velocity exceeds the maximum group velocity $V_0$, the soliton leaves behind a "tail" of the chain oscillations which leads to a finite path length of the soliton. For $(V-V_0)/V_0 \ll 1$, the path length will be:

$$L\big/a \sim \exp \frac{\pi\, r\, a\, \Omega_0}{\sqrt{V^2 - V_0^2}} \qquad (26)$$

As $V_0 = 0$ expression (26) transforms into (25).

In another limit case of large velocities ($\pi\, r\, a\, \Omega_0 / 2V \ll 1$), when

$$V \gg 2\pi \frac{M}{m} (\hbar \Omega_0)^2 \frac{\Omega_0}{\chi^2 a} > V_0 \qquad (27)$$

the soliton path length is written (20)-(24) as:

$$L\big/a = \frac{1}{30}\left(\frac{m}{M}\right)^3 \frac{a^2 \chi^6 V^4}{\Omega_0^{10} \hbar^6} \qquad (28)$$

So, at large $V$ the path length grows as $\sim V^4$. The physical reason is that in the case of $V \to \infty$ from Cherenkov condition it follows that $q \to 0$. As $q \to 0$ a portion of electron distribution taking part in the phonon emission is independent of $V$. Accordingly, the soliton path length will tend to infinity as $V \to \infty$.

**$\varepsilon > 0$.** When $\varepsilon > 0$, the soliton velocity is less than the maximum group velocity of the chain $V_0$. Using (17), (12) we express $u(\xi)$ as:

$$u(\xi) = -\frac{1}{2\sqrt{|\varepsilon|}} \frac{\chi}{M \Omega_0^2} \int d\xi' e^{-|\xi'-\xi|/\sqrt{|\varepsilon|}} |\varphi(\xi')|^2 \qquad (29)$$

From (29) follows that in a steady case the displacements propagate both behind the moving soliton and in front of it, dying out exponentially with distance from the soliton. In this case the quantity $E_{ph}$ has a finite value even if the limits of integration in integral (24) are infinite and corresponds to the energy of the phonon environment accompanying the charge motion along the chain. Then the path length will be infinite.

## 5. Soliton path length in the presence of dissipation. Soliton mobility.

In the presence of dissipation the soliton path length always has a finite value. As $\delta \neq 0$ the loss of energy by a soliton $dH/dt$ due to dissipation is determined by the balance energy equation:

$$\frac{dH}{dt} = -2F, \qquad F = \frac{\gamma}{2a} \int \left(\frac{\partial u}{\partial t}\right)^2 dx, \qquad (30)$$

where $F$ is a dissipative function, $\gamma = M\Omega'$ is a friction coefficient. From (8), (30) follows:

$$\frac{dH}{dt} = -\frac{M\Omega'}{a^2} V^2 \int \left(\frac{du}{d\xi}\right)^2 d\xi \qquad (31)$$

When the soliton moves at a small velocity, its energy can be presented as [19]:

$$H = H_0 + \frac{1}{2} m^{**} V^2, \qquad (32)$$

$$m^{**} = \frac{M}{a^2} \int \left(\frac{du}{d\xi}\right)^2 d\xi, \qquad (33)$$

where $m^{**}$ is the soliton effective mass. With the use of (32), (33) balance equation (31) takes the form:

$$\frac{dV^2}{dt} = -2\Omega' V^2 \qquad (34)$$

From (34) follows the expression for the soliton velocity:

$$V(t) = V(0) e^{-\Omega' t} \qquad (35)$$

where $V(0)$ is the soliton initial velocity. From (35), the soliton relaxation time will be $\tau_r = 1/\Omega'$. Using (35) we express the soliton path length as:

$$L = \int_0^\infty V(t) dt = V(0)/\Omega' \qquad (36)$$

Now let us discuss the motion of a soliton at a small velocity in an external electric field $\mathcal{E}$. Given an electric field, in the steady case the loss of energy due to dissipation ($\dot{H} = -2F$) will be counterbalanced by a gain in energy obtained by the soliton from the external field ($\dot{H} = e\mathcal{E}V$). As a result, the balance energy will take the form:

$$e\,\mathcal{E}\,V = \Omega' m^{**} V^2 \tag{37}$$

Whence the soliton mobility $\mu$ is:

$$\mu = e\,\tau_r\,/\,m^{**} \tag{38}$$

This is the ordinary expression for the electron mobility where effective mass of soliton stands in place of the effective mass of electron and the relaxation time is $\tau_r = 1/\Omega'$. We will turn back to this point in § 7, 8.

Notice that according to (38), the soliton mobility is determined only by dissipation which leads to stopping of the soliton. According to the analysis carried out above, the contribution of phonon emission into the mobility vanishes in the limit considered $V \to 0$. Below we will take the case of a finite soliton velocity when the contribution of emission into its value plays a decisive role.

# 6. Motion in an electric field.

To calculate the equilibrium velocity of a soliton in an external electric field of intensity $\mathcal{E}$ we will proceed from the energy dissipation expression (30). In the steady case the rate of energy dissipation is counterbalanced by the energy gained by the particle moving in an electric field $\mathcal{E}$ at the velocity $V$:

$$-\frac{dH}{dt} = e\mathcal{E}V \tag{39}$$

Expanding $|\varphi(\xi)|^2$ into Fourier series:

$$\left|\varphi\left(\xi\right)\right|^2 = N^{-1} \sum_k e^{ik\xi} C_k \tag{40}$$

and using relations (12) –(14) and (39) – (40), we express the energy dissipation rate as:

$$\frac{dH}{dt} = \frac{\gamma}{2\pi a^2} \frac{\chi^2}{\left(M\,\Omega_0^2\right)^2} \int \frac{k^2 V^2 C_k C_{-k}}{\left(\varepsilon k^2 + 2i\,\delta k + 1\right)\left(\varepsilon k^2 - 2i\,\delta k + 1\right)}\,dk \tag{41}$$

From (19), (38) we get:

$$C_k = \frac{\pi}{2} \frac{k\,r}{\mathrm{sh}\left(\pi k\,r/2\right)} \tag{42}$$

Using relations (37), (41), (42) we obtain the following general expressions for the dependence of the soliton velocity on the electric field intensity:

1) when $\omega' < 2\omega$, $v > v_0 / \sqrt{1 - (\omega'/2\omega)^2}$ :

$$E = 2\,\omega^2\,\omega'\,\frac{\eta v}{\left(v_0^2 - v^2\right)^2} \int\limits_0^\infty \frac{x^4 \operatorname{sh}^{-2} x}{\left(x^2 + c_1\right)^2 + c_2^2}\,dx\,, \qquad (43)$$

$$c_1 = 4\pi^2\,\frac{\eta^2}{\kappa^2}\,\frac{\omega^2}{\left(v_0^2 - v^2\right)}\left(1 + \frac{\omega'^2}{2\omega^2}\,\frac{v^2}{\left(v_0^2 - v^2\right)}\right),$$

$$c_2^2 = \left(\frac{2\pi\eta}{\kappa v}\right)^4\,\frac{v^6 \omega'^2}{\left(v^2 - v_0^2\right)^3}\left(\omega^2 + \left(\frac{\omega'}{2}\right)^2\,\frac{v^2}{\left(v_0^2 - v^2\right)}\right);$$

2) when $\omega' < 2\omega$, $v < v_0 / \sqrt{1 - (\omega'/2\omega)^2}$ :

$$E = 2\,\omega^2\,\omega'\,\frac{\eta v}{\left(v_0^2 - v^2\right)^2} \int\limits_0^\infty \frac{x^4 \operatorname{sh}^{-2} x}{\left(x^2 + c_1^2\right)\left(x^2 + c_2^2\right)}\,dx\,, \qquad (44)$$

$$c_1^2 = \left(\frac{2\pi\eta}{\kappa}\right)\frac{\omega^2}{v_0^2 - v^2}\left[1 + \frac{\omega'^2}{2\omega^2}\,\frac{v^2}{v_0^2 - v^2} + \frac{v\omega'/\omega^2}{\sqrt{v_0^2 - v^2}}\sqrt{\omega^2 + \left(\frac{\omega'}{2}\right)^2\,\frac{v^2}{v_0^2 - v^2}}\right]$$

$$c_2^2 = \left(\frac{2\pi\eta}{\kappa}\right)\frac{\omega^2}{v_0^2 - v^2}\left[1 + \frac{\omega'^2}{2\omega^2}\,\frac{v^2}{v_0^2 - v^2} - \frac{v\omega'/\omega^2}{\sqrt{v_0^2 - v^2}}\sqrt{\omega^2 + \left(\frac{\omega'}{2}\right)^2\,\frac{v^2}{v_0^2 - v^2}}\right]$$

3) when $\omega' > 2\omega$: for any $v$ the dependence $E(v)$ is determined by the formulae of case 2).

In expressions (43), (44) use is made of dimensionless quantities: $E$, $\eta$, $\omega$, $\omega'$, $\kappa$, $v$, which are related to the dimensional quantities as:

$$E = \mathcal{E}\,\frac{e\,a\,\tau}{\hbar}\,, \qquad \eta = \frac{\hbar\tau}{2m\,a^2}\,, \qquad \omega = \Omega_0\,\tau\,, \qquad \omega' = \Omega'\,\tau,$$

$$\kappa\omega^2 = \tau^3\chi^2 / M\hbar\,, \qquad v = V\tau/a\,, \qquad v_0 = V_0\tau/a\,, \qquad (45)$$

where $\tau$ is an arbitrary time scale.

Recall that expressions (43), (44) are not the exact solution of the problem (9), (10) for all the parameter values. The reason is that when deriving (43), (44) we used approximate expression (19) and its corresponding expression (42). Calculation experiments demonstrate that expression (19) is still a good approximation even at $r\sim1$, when the chain discreteness is essential [26]. Up to rather high velocities, the soliton shape turns out to be close to that described by (19). So, we may hope that relations (43), (44) are valid within a wide range of the model parameters variation. At the same time it should be noted that (43), (44) are invalid, for example, in the limit case $\Omega_0=0$ which corresponds to the limit $|\varepsilon|\to\infty$. It is easy to show that any localized soliton states are lacking in this case (acoustic phonon spectrum) and the solution of (10), (29) will be delocalized electron states. Hence, taking account of dispersion for $\Omega_0\neq0$ leads to enlargement of the soliton, i.e. deviation of its shape from that given by (19). Notice that as $v = v_0$, nothing unusual peculiarity arises, in particular, the electric field intensity necessary to impart this velocity to the soliton, according to (43), (44), is equal to:

$$E = \frac{\pi^2}{60}\frac{\kappa^4 v_0^2 \omega'}{\omega^2 \eta^3}$$

In the limit case $v_0=0$, i.e. when dispersion is absent, for cases 1), 3) we get from (43), (44) the following expressions:

$$E = \kappa\,\omega^2\,\omega'\,v\,I \tag{46}$$

$$I = \frac{2\eta}{\kappa v^4}\int\limits_0^\infty \frac{x^4/\mathrm{sh}^2 x}{\left(x^2 + c_1\right) + c_2^2}\,dx\,, \tag{47}$$

$$c_1 = \left(\frac{2\pi\eta}{\kappa v}\right)^2\left(\frac{\omega'^2}{2} - \omega^2\right),$$

$$c_2^2 = \left(\frac{2\pi\eta}{\kappa v}\right)^4 \omega'^2\left(\omega^2 - \left(\frac{\omega'}{2}\right)^2\right),$$

for $\omega > \omega'/2$. In the case of a damped motion, when friction is large, we get the following expression of $I$:

$$I = \frac{2\eta}{\kappa v^4}\int\limits_0^\infty \frac{x^4/\mathrm{sh}^2 x}{\left(x^2 + c_1^2\right)\left(x^2 + c_2^2\right)}\,dx\,, \tag{48}$$

$$c_1 = \frac{2\pi\eta}{\kappa v}\left(\frac{\omega'}{2} - \sqrt{\left(\frac{\omega'}{2}\right)^2 - \omega^2}\right),$$

$$c_2 = \frac{2\pi\eta}{\kappa v}\left(\frac{\omega'^2}{2} + \sqrt{\left(\frac{\omega'}{2}\right)^2 - \omega^2}\right),$$

for $\omega' > 2\,\omega$.

Let us consider the limit case of the lack of friction $\omega' = 0$. From (46), (47) it follows:

$$E = 2\pi^2 \frac{\omega^2 \eta^2}{\kappa} \frac{1}{v^4} \frac{1}{\operatorname{sh}^2(2\pi\eta\omega/\kappa v)}, \qquad (49)$$

So, according to (49), even in the absence of friction, for a soliton to move steadily along the chain, a nonzero field $E$ should be applied. This result is in full agreement with the analysis carried out above. As $\omega' = 0$, the work of an electric field is necessary for a moving soliton to induce oscillations in the chain.

Fig.1 shows dependencies of the soliton velocity on the intensity of the electric field $E$ determined by (46) - (48) for various values of the parameter $\omega'$ in the absence of dispersion.

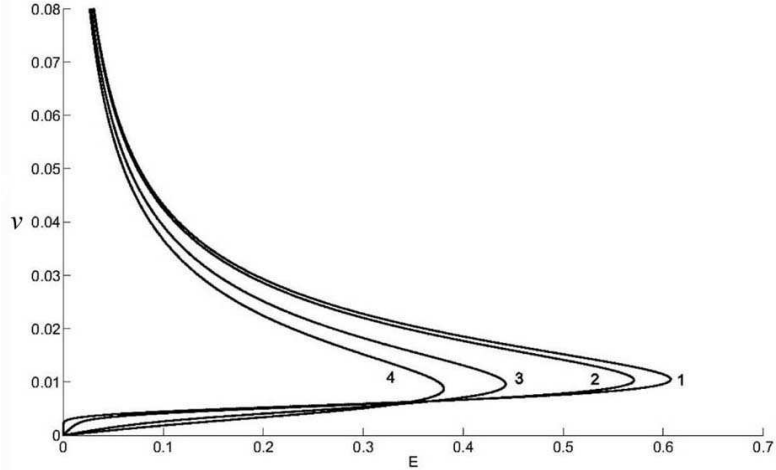

Figure 1. The dependence of soliton velocity $v$ on electric fild $E$ for different values of parameter $\omega'$: 1)$\omega'$=0; 2)$\omega'$=0.001; 3)$\omega'$=0.006 4)$\omega'$=0.01; ($\kappa$=4, $\eta$=1.276, $\omega$=0.01).

The limit case of $\omega' = 0$, determined by (49) corresponds to curve 1 in Fig.1 and the limit case of $\omega' = \infty$ corresponds to the ordinate axis. As is seen from Fig.1, a steady motion of a soliton is possible only within the interval $0 < E < E_{max}(\omega')$. On each curve, the branch for which $dv/dE = v'_E > 0$ corresponds to a steady motion of a soliton, while the branch for which $v'_E < 0$ corresponds to its unsteady motion.

From (46)-(48) it follows that the occurrence of friction ($\omega' \neq 0$) gives rise to an ohmic branch, i. e. a linear dependence of $v(E)$ (for small $E$) on the stable regions of $v(E)$ curves.

# 7. Soliton mobility for T=0.

At zero temperature and $v_0 \neq 0$, (44) yields the following expression for the soliton mobility:

$$\mu = \frac{e\,a^2}{\hbar} \frac{v_0^4}{2\omega^2 \omega' \eta I_b}, \qquad (50)$$

where

$$I_b = \int_0^\infty \frac{x^4/\operatorname{sh}^2 x}{\left(x^2 + b^2\right)^2}\, dx, \qquad b = \frac{2\pi\eta\omega}{\kappa v_0} \qquad (51)$$

From (50), (51) it follows that in the limit $v_0 \rightarrow \infty$ :

$$\mu = 4\frac{e\,a^2}{\hbar}\,\frac{v_0^3}{\omega\omega'\kappa}, \tag{52}$$

As $v_0 = 0$, from (44) we get the following expression for the soliton mobility:

$$\mu = 240\frac{e\,a^2}{\hbar}\,\frac{\eta^3\,\omega^3}{\kappa^4\omega'}, \tag{53}$$

which coincides with earlier obtained expression (38).

Fig.2 shows in greater detail linear portions of the $v(E)$ dependencies corresponding to curves in fig.1.

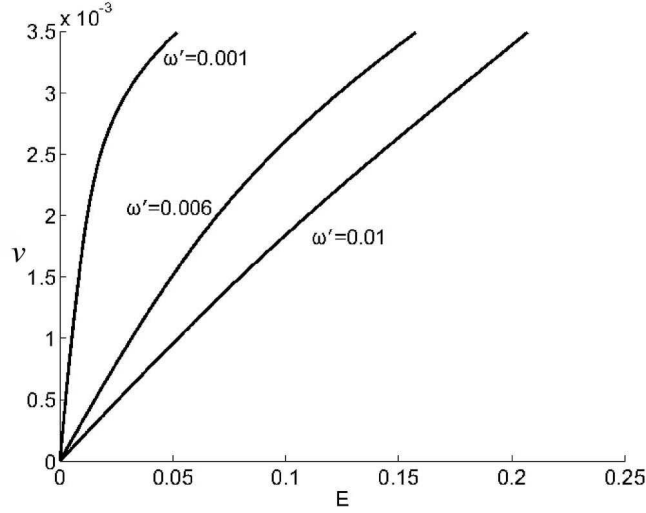

Figure 2. Linear portions of $v\,(E)$ curves.

Fig.2 suggests that the portions of linear dependence lengthen as $\omega'$ grows. For $\omega' \geq 0{,}1$ the regions of $v(E)$ linear dependence spread up to the critical value of the field intensity at which a steady motion of a soliton becomes impossible. In this case Ohm's law holds up to very high values of the field intensity.

To understand the meaning of expressions (50) – (53) for the soliton mobility let us calculate the soliton effective mass $m^{**}$ with the use of (33). When from $v_0 = 0$ (33) we get:

$$m^{**}\big/m = \left(1\big/120\right)\left(\kappa^4\big/\eta^2\,\omega^2\right) \tag{54}$$

Substitution of (54) into (38) yields expression (53). So, the soliton mobility is described by an ordinary equation in which the effective mass is replaced by the soliton mass and the relaxation time – by the quantity $1/\Omega'$. Hence, for $V_0 = 0$, the soliton effective mass is equal to:

$$m^{**}\big/m = 4\,\eta^2\,\omega^2\,I_b\big/v_0^4 \tag{55}$$

Really the obtained results referes to the temperature which is considerably less than that of soliton decomposition [27].

## 8. Application to polynucleotide chain.

Recently, modeling of the motion of a charged particle in molecular chains of various types has been the topic of considerable investigation (books and reviews [9, 28-31] and papers cited therein). A new type of conducting quasi-one-dimensional system is polynucleotide chains inter twisted into a double helix of DNA molecule. There are a lot of works where a problem of soliton or polaron motion in DNA has been studied, [13, 31 - 37].

Recall, that DNA consists of four types of nucleotides designated as A (adenine), T(thymine), C(cytosine), an G(guanine) which unite into complementary pairs in such a way that nucleotide A always pairs with T and nucleotide C always pairs with G. These nucleotide pairs are arranged in a stack to form a DNA double helix. Now days long sequences with a prescribed set of nucleotide pairs can be synthesized artificially. Of great interest are chains composed of uniform pairs which could serve as molecular wires in nanoelectronic devices [1]. In the majority of experiments on charge transfer in DNA the charge is carried not by electrons, but by holes. If a nucleotide in nucleotide chain be freed from electron, the hole which would arise would have a potential energy $U$ such that $U_G < U_A < U_C < U_T$.

Overlapping of electron $\pi$ - orbitals of neighboring pairs will lead to delocalization of the hole over the chain and its capture by nucleotides with lower oxidation potential. Since, according to the above inequality, guanine has the lowest oxidation potential, the hole will travel over guanines while all the other nucleotides will act as potential barriers.

In our approach Hamiltonian of a regular deformable nucleotide chain is considered in continuum approximation. To carry out numerical estimates we choose the following parameter values. For a homogeneous PolyG/PolyC chain the value of the matrix element $v_{GG}$ is equal to 0,084 eV which corresponds to $\eta = 1,276$. The values of matrix elements for DNA calculated by quantum chemical methods in [38, 39], are in good agreement with experimental data on relative charge transfer rates in DNA. The values $\chi = 0,13$ eV/Å, $\Omega_0 = \sqrt{K/M} = 10^{12}$ sec$^{-1}$, $\Omega' = \gamma/M$ = 6 10$^{11}$ sec$^{-1}$ are chosen the same as in [14]. Notice, that the value of $\chi$ used by us is close to that found by quantum chemical methods in [40]: $\chi = 0,2349$ eV/Å. For $\tau = 10^{-14}$ sec these parameter values correspond to dimensionless quantities $\kappa = 4$, $\omega = 10^{-2}$, $\omega' = 0,006$. The value of mobility $\mu$ calculated by formula (53) for the distance between nucleotide pairs $a = 3,4$Å in a PolyG/PolyC chain is $\mu \approx 5,7 \ 10^{-2}$ cm$^2$/V sec. The characteristic size of a soliton is $r \approx 1,3$, which corresponds to the energy of the particle $W_0$ reckoned from the bottom of the conduction band: $W_0 = -\hbar /2 \ m^* r^2 \approx -5 \ 10^{-2}$ eV and the soliton energy $E_0 = W_0 / 3 = -1,7 \ 10^{-2}$ eV. The soliton effective mass calculated by formula (54) is $m^{**} \approx 1,3 \ 10^4 \ m^*$ where the band mass of a hole $m^* = -\hbar /2 \ v \ a^2$ is equal to 3,94 $m_0$, $m_0$ is the electron mass in vacuum.

## 9. Discussion of results.

In Bogolubov-Tyablikov strong coupling theory [41, 42] which is presently considered the most consistent theory of a strong coupling polaron (see, for example, review [43]), at $V \neq 0$ some resonance denominators appear which lead to divergence. The results obtained here provide an explanation for this contradiction. In the case under consideration we could have proceeded from Froehlich Hamiltonian for the Holstein model and get resonance denominators (see, for example, discussion of this problem in [31]). So, the presence of resonance denominators stems from the fact that a moving polaron state is nonstationary, i.e. it is not the eigen state of Froehlich Hamiltonian. This implies an important methodological conclusion – in translation-invariant systems, among which is Froehlich Hamiltonian, a moving polaron state may not be its eigen state.

In a semiclassical model in a dispersionless case, a moving polaron always emits phonons. In a quantum case this emission occurs on the condition $m^{**}v^2/2 > \hbar\Omega_0$. One would think that in a quantum case, as distinct from the classical one, there exists a mimimum velocity $v_{\min} = 2\hbar\Omega_0/m^{**}$, at which the emission is possible. However, the very inequality from which $v_{\min}$ was obtained is fulfilled only in the strong coupling limit, when quantum description becomes semiclassical. Since in the strong coupling limit $m^{**} \sim \kappa^4$, where $\kappa \to \infty$, then $v_{\min} \sim 1/\kappa^4 \to 0$ in complete agreement with a semiclassical approach.

The general picture of the soliton motion along the chain at $T=0$ in the absence of dissipation looks as follows. For $r \gg 1$, a soliton, irrespective of the model parameters, has a large path length which goes into infinity as $v \to 0$ and $v \to \infty$. In the presence of dispersion when $v < v_0$ the soliton path length is $L = \infty$. When $v < v_0$ the path length is finite and tends to infinity as $v \to \infty$.

We emphasize that despite the fact that the value of path length $L$ at given $V > V_0$ is finite, the total soliton path for a soliton to reach the velocity $V_0$ is always infinite. It is clear that as the soliton is slowing down, i.e. its velocity decreases, the local amplitude of the phonone tail vanishing at $V \to V_0$. Accordingly the soliton path length tends to infinity at $V \to V_0$ and an infinite time is required for a soliton to reach the velocity $V_0$.

When $r \ll 1$, a soliton cannot move along the chain. The deformation produced by it "chains' it to the site (molecule) where it is localized (when considered discretely). Formally, in the continuum model, which is invalid in this case, the soliton path length becomes very small: $L \le r$. The reason why the deformation "chains" the charge to the site in the case of interest, is that the ratio of the deformation energy $E_{pot}$ to the oscillator energy: $E_{pot}/\hbar\Omega_0 = \kappa/2\omega$ is very large: $\kappa/2\omega \gg 1$ and in passing to the neighboring site the charge must do some work so that to produce an equilibrium deformation at this site. For $E_{kin} < E_{pot}$ this process is forbidden by the energy conservation law: when $E_{pot} = \chi^2/2M\Omega_o^2 > m^{**}V^2/2$ a soliton cannot move to the neighboring site.

In the quantum model, when $r \le 1$, the description is carried out in terms of a small radius polaron (SRP). Agreement with the semiclassical description is achieved, as in the case of a large radius soliton, by passing to the limit $\kappa \to \infty$ in the equation $v_{\min} = 2\hbar\Omega_0/m^{**}$, where $m^{**}$ - is SRP effective mass.

So, for a SRP $v_{\min} \to 0$ as $\kappa \to \infty$. It follows that at $T = 0$ the polaron band, the concept of which was first introduced for a SRP by Tyablikov [44], does not exist. As was pointed out above, the translation invariance of the Hamiltonian, on which Tyablikov's theory relies, does not guarantee the existence of the band.

From all has been said it follows that any experimental attempts to find the band of a small radius polaron (SRP) at $T = 0$ have no prospects. Traditionally, the difficulty in finding the SRP band is associated with its narrowness when any external excitation leads to its destroy and formation of a localized state (see discussion of this problem in [45]). According to our results, even in the absence of any external effects the band will destroy by a phonon emission of a moving polaron which leads to its immediate localization.

The work was supported by RFBR, project N 07-07-00313.